\documentclass{webofc}
\usepackage[varg]{txfonts}   
\usepackage{epsfig}
\usepackage{graphicx}
\usepackage{mathptmx}      
\usepackage{bm}
\newcommand{\BG}{\,{\textrm{\tiny BG}}}
\newcommand{\TF}{\,{\textrm{\tiny TF}}}

\newcommand{\GeV}{\,{\mathrm{GeV}}}

\begin{document}
\title{Fractal structure of hadrons and non-extensive statistics~\thanks{Presented by E.~Meg\'{\i}as at the QCD@Work: International Workshop on QCD, 25-28 June 2018, Matera, Italy.}}
%
%

\author{\firstname{Eugenio} \lastname{Meg\'{\i}as}\inst{1,2}\fnsep\thanks{\email{emegias@ugr.es}} \and
        \firstname{Airton} \lastname{Deppman}\inst{3}\fnsep\thanks{\email{deppman@if.usp.br}} \and
        \firstname{Tobias} \lastname{Frederico}\inst{4}\fnsep\thanks{\email{tobias@ita.br}} \and
        \firstname{D\'ebora P.} \lastname{Menezes}\inst{5}\fnsep\thanks{\email{debora.p.m.26@gmail.com}}
}

\institute{Departamento de F\'{\i}sica At\'omica, Molecular y Nuclear and Instituto Carlos I de F\'{\i}sica Te\'orica y Computacional, Universidad de Granada, Avenida de Fuente Nueva s/n, 18071 Granada, Spain
\and
Departmento de F\'{\i}sica Te\'orica, Universidad del Pa\'{\i}s Vasco UPV/EHU, Apartado 644, 48080 Bilbao, Spain
\and
    Instituto de F\'{\i}sica, Universidade de S\~ao Paulo, Rua do Mat\~ao Travessa R Nr. 187, Cidade Universit\'aria, CEP 05508-090 S\~ao Paulo, Brazil
\and
    Instituto Tecnol\'ogico da Aeron\'autica, 12228-900 S\~ao Jos\'e dos Campos, Brazil
\and
    Departamento de F\'{\i}sica, CFM, Universidade Federal de Santa Catarina, CP 476, CEP 88040-900 Florian\'opolis, Brazil
          }

\abstract{The role played by non-extensive thermodynamics in physical
  systems has been under intense debate for the last decades. Some
  possible mechanisms that could give rise to non-extensive statistics
  have been formulated along the last few years, in particular the
  existence of a fractal structure in thermodynamic functions for
  hadronic systems. We investigate the properties of such fractal
  thermodynamical systems, in particular the fractal scale invariance
  is discussed in terms of the Callan-Symanzik~equation. Finally, we
  propose a diagrammatic method for calculations of relevant
  quantities.}
\maketitle
\section{Introduction}
\label{intro}

Entropy emerges as an important quantity in different areas that try to describe systems of increasing complexity. The formulation of new entropic forms that generalize the one proposed by Boltzmann constitutes an important research line. In particular, the non-additive entropy introduced by Tsallis~\cite{Tsallis:1987eu} has found wide applicability in the last few years, see e.g.~\cite{Tempesta:2011vc,Kalogeropoulos:2014mka}. However, the full understanding of the non-extensive statistics formulated by Tsallis has not been accomplished yet. Several connections between Boltzmann and Tsallis statistics have been proposed so far, see e.g.~\cite{Borland:1998,Wilk:2009nn,Deppman:2016fxs}, but it seems that the physical meaning of the entropic index~$q$ is not understood in the general case. In the present work, we make a detailed analysis of the connection based on a system featuring fractal structure in its thermodynamic properties. Fractals are conceived as objects with an internal structure that can be considered as an ideal gas of a specific number of subsystems, which, in turn, are also fractals of the same kind. The self-similarity between fractals at different levels of the internal structure reveals the typical scale invariance. The results obtained in the present work offer a new perspective in the analysis of hadron structure.

\section{Tsallis statistics and QCD thermodynamics}
\label{Tsallis}

Tsallis statistics constitutes a generalization of Boltzmann-Gibbs (BG) statistics, under the assumption that the entropy of the system is non-additive. For two independent systems $A$ and $B$
\begin{equation}
S_{A+B} = S_A + S_B + (1-q)S_A S_B \,,
\end{equation}
where the entropic index $q$ measures the degree of non-extensivity~\cite{Tsallis:1987eu}. Let us define the $q$-exponential $e_q^{(\pm)}(x)=[1 \pm (q-1)x]^{\pm1/(q-1)}\,$, with $e_q^{(+)}(x)$ defined for $x\ge 0$ and $e_q^{(-)}(x)$ for $x < 0$, and the $q$-log function $\log^{(\pm)}_q(x)=\pm (x^{\pm(q-1)}-1)/(q-1) \,$. Then the grand-canonical partition function for a non-extensive ideal quantum gas at finite chemical potential is~\cite{Megias:2015fra}
\begin{eqnarray}
 \log\Xi_q(V,T,\mu) &=&
 -\xi V\int \frac{d^3p}{{(2\pi)^3}} \sum_{r=\pm}\Theta(r x)\log^{(-r)}_q\bigg(\frac{ e_q^{(r)}(x)-\xi}{ e_q^{(r)}(x)}\bigg) \,, \label{partitionfunc}
\end{eqnarray}
where $x = (E_p - \mu) / (kT)$, the particle energy is $E_p = \sqrt{p^2+m^2}$ with $m$ the mass and $\mu$ the chemical potential, $\xi = \pm 1$ for bosons and fermions respectively, and $\Theta$ is the step function. Eq.~(\ref{partitionfunc}) reduces to the Bose-Einstein and Fermi-Dirac partition functions in the limit $q \rightarrow 1$. 

The thermodynamics of Quantum Chromodynamics (QCD) in the confined phase can be studied within the Hadron Resonance Gas (HRG) approach, which is based on the assumption that physical observables in this phase admit a representation in terms of hadronic states which are treated as non-interacting and point-like particles~\cite{Hagedorn:1984hz}. These states are taken as the conventional hadrons listed in the review by the Particle Data Group. Within this approach the partition function is then given by~\cite{Megias:2015fra,Menezes:2014wqa}
\begin{equation}
\log \Xi_q(V,T,\{\mu\})=\sum_i \log \Xi_q(V,T,\mu_i)\,, \label{eq:logZ}
\end{equation}
where $\mu_i$ refers to the chemical potential for the {\it i-th} hadron. 
\begin{figure*}[htb]
\centering
\includegraphics[width=5.7cm]{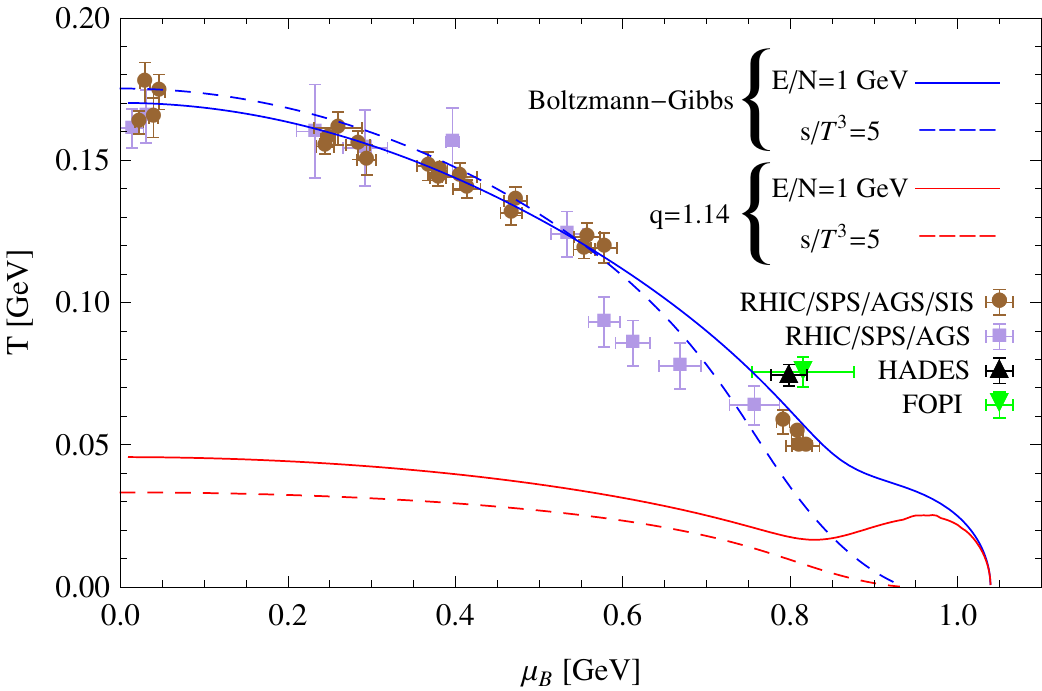} \hspace{1cm}
\includegraphics[width=5.7cm]{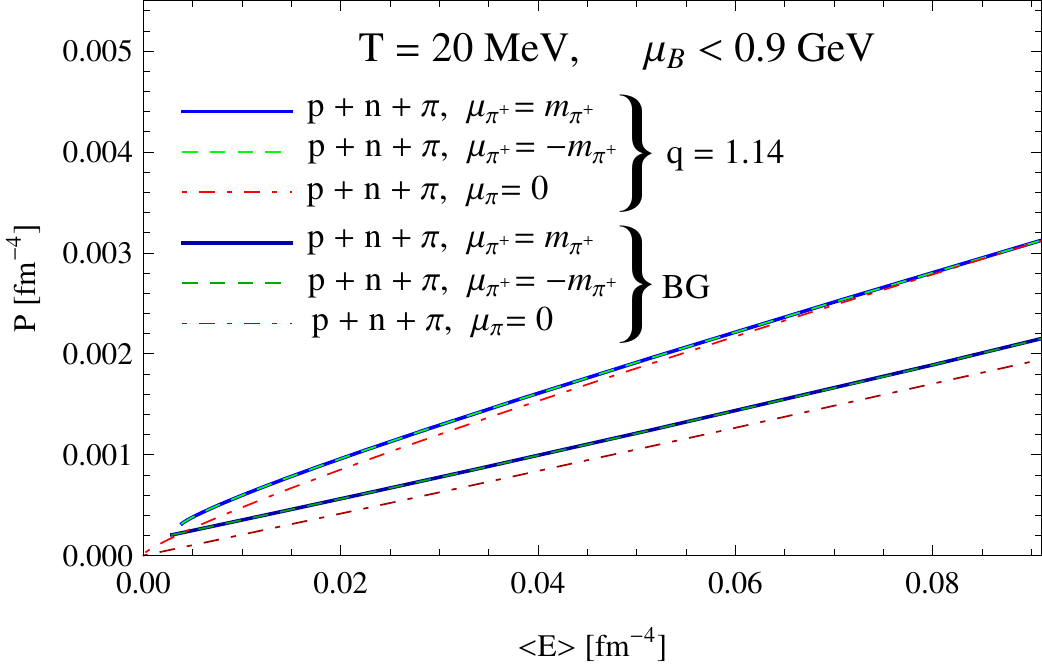} 
\caption{Left panel: Chemical~freeze-out~line~$T = T(\mu_B)$. Right panel: EoS of hadronic matter.}
\label{fig:1}
\end{figure*}
The thermodynamic functions can be obtained from Eq.~(\ref{eq:logZ})
by using the standard thermodynamic relations. There are different
proposals for the conditions determining the transition line between
the confined and the deconfined regimes of QCD. Using the arguments
of~\cite{Cleymans:1999st}, the phase transition line in the $T \times
\mu_B$ diagram can be determined by the condition $\langle E\rangle /
\langle N \rangle = 1 \GeV$. A different method based on the entropy
density has been proposed in~\cite{Tawfik:2005qn}. The result for the
chemical freeze-out line is displayed in Fig.~\ref{fig:1} (left). We
also display in Fig.~\ref{fig:1} (right) the Equation of State (EoS)
for hadronic matter at finite baryonic chemical potential. It it
remarkable that $P(E)$ becomes larger in Tsallis statistics as
compared to BG statistics (see a discussion
in~\cite{Megias:2015fra,Menezes:2014wqa}).

\section{Thermofractals and Tsallis statistics}
\label{sec:Thermofractals}

The emergence of the non-extensive behavior has been attributed to
different causes: long-range interactions, correlations and memory
effects~\cite{Borland:1998}; temperature
fluctuations~\cite{Wilk:2009nn}; and finite size of the system. In
this work we will study a natural derivation of non-extensive
statistics in terms of thermofractals. These are systems in
thermodynamical equilibrium presenting the following
properties~\cite{Deppman:2016fxs,Deppman:2016prl}:
\begin{enumerate}
\item The total energy is given by
\begin{equation}
U = F + E \,,
\end{equation}
where $F$ is the kinetic energy of $N$ constituent subsystems, and $E$ is the internal energy of those subsystems, which behave as particles with an internal structure.
\item The constituent particles are thermofractals. The energy distribution $P_{\TF}(E)$ is self-similar or self-affine, so that at some level $n$ of the hierarchy of subsystems, $P_{\TF (n)}(E)$ is equal to the distribution in the other levels.
\item At level $n$ in the hierarchy of subsystems the phase space is so narrow that one can consider~$P_{\TF (n)}(E_n) dE_n = \rho dE_n$, with $\rho$ being independent of the energy~$E_n$.
\end{enumerate}

The energy distribution of a thermodynamical system is given, according to BG statistics, by
\begin{equation}
 P_{\BG}(U) dU=A \exp(-U/kT) dU\,, \label{Uvar}
\end{equation}
where $A$ is a normalization constant. However, in the case of thermofractals the phase space must include the momentum degrees of freedom of free particles as well as their internal degrees of freedom. According to property 2 of self-similar thermofractals, the internal energy is given by~\cite{Deppman:2016fxs}
\begin{equation}
dE = \frac{F}{kT} \left[ P_{\TF}(\varepsilon) \right]^\nu d\varepsilon \,,
\end{equation}
where~$\varepsilon/(kT) = E/F$ and $\nu$ is to be determined. Then, the total energy distribution is given by
\begin{equation}
P_{\TF (0)}(U) dU = A^\prime F^{\frac{3N}{2}-1} \exp\left( -\frac{\alpha F}{kT} \right) dF  \left[ P_{\TF (1)}(\varepsilon) \right]^\nu d\varepsilon \,, \qquad \alpha = 1 + \frac{\varepsilon}{kT} \,.
\end{equation} 
This expression relates the distributions at level $0$ and level $1$ of the subsystem hierarchy. After integration in $F$, one finds
\begin{equation}
 \Omega = \int P_{\TF (0)}(U) dU = \int_0^{\infty} A \cdot \bigg[1+\frac{\varepsilon}{kT}\bigg]^{-3N/2} \left[P_{\TF (1)}(\varepsilon)\right]^{\nu} d\varepsilon \,, \qquad A = \Gamma\left[\frac{3}{2}N\right] (kT)^{\frac{3}{2}N} A^\prime \,,  \label{Oeps}
\end{equation}
where $\Gamma[z]$ is the Euler gamma function, and obviously $\Omega = 1$. At this point it is possible to impose the identity
\begin{equation}
 P_{\TF (0)}(U) \propto P_{\TF (1)}(\varepsilon)\,, \label{similarity}
\end{equation}
corresponding to a self-similar solution for the thermofractal probability distribution. Then, the simultaneous solution of Eqs.~(\ref{Oeps}) and~(\ref{similarity}) is obtained with~\cite{Deppman:2016prl,Deppman:2017fkq}
\begin{equation}
P_{\TF (n)}(\varepsilon)= A_n \cdot \bigg[1 + (q-1)\frac{\varepsilon}{k\tau}\bigg]^{-\frac{1}{q-1}}   \,, \label{selfsimilar}
\end{equation}
where~$q-1 = 2(1-\nu)/(3N)$ and $\tau = N(q-1) T$. The energy distribution of thermofractals then obeys Tsallis statistics, and this constitutes one of the most important results of the present work.~\footnote{In the following we will define $N^\prime = N + 2/3$ as the effective number of particles taking into account their internal degrees of freedom (see~\cite{Deppman:2017fkq} for details).}

\section{Callan-Symanzik equation and diagrammatic representation}
\label{sec:Diagram_CS}

Due to the evident similarities between hadron structure and thermofractal structure~\cite{Deppman:2016prl,Deppman:2012us,Megias:2014tha,Deppman:2015cda,Deppman:2017igr}, it is possible to show that the thermofractal description has close connections to quantum field theory as far as scaling properties are concerned. This will be done in a future work~\cite{Deppman:inprep}, but we will advance some aspects as follows.

\subsection{Scale invariance and Callan-Symanzik equation}
\label{subsec:CS_eq}

Thermofractals are scale invariant. Formally, this property should be accomplished with the scale invariance of the distribution of kinetic and internal energy. Using $F^{(0)}/T^{(0)} = F^{(n)}/T^{(n)}$ and $\varepsilon/(kT) = E^{(n)}/F^{(n)}$, one finds~\cite{Deppman:2017fkq}
\begin{equation}
 \lambda_n := \frac{E^{(n)}}{E^{(0)}} = \left( \frac{1}{N} \right)^{\frac{n}{1-D}} \,,
\end{equation}
where $D$ is the fractal dimension. Using thermofractals it can be obtained the fractal dimension of hadrons, resulting in $D = 0.69$~\cite{Deppman:2016fxs}, a value close to that resulting from intermittence analysis~\cite{Ghosh:1998gw}.  Using these scaling properties, the vertex function of thermofractals writes~\cite{Deppman:2017fkq}
\begin{equation}
\Gamma(E,\varepsilon,T) \propto (kT^\prime)^{-(1-D)} g  \prod_{i=1}^{N^\prime} \left( 2\pi \frac{E_i}{k T^\prime} \right)^{-3/2}  \left[ P_{\TF}(\varepsilon_i) \right]^\nu  \,,
\end{equation}
where $T^\prime = T^{(n)}$. Introducing $M = k T^\prime$ as the scale parameter that determines the fractal structure of the subsystem at a certain level, then one can derive the Callan-Symanzik equation for thermofractals, which writes
\begin{equation}
\left[ M \frac{\partial}{\partial M} + \sum_{i=1}^{N^\prime} E_i \frac{\partial}{\partial E_i} + d \right] \Gamma = 0 \,,
\end{equation}
where $d = 1 - D$ is the anomalous dimension for thermofractals. This result is equivalent to the one obtained in~\cite{Deppman:2017rtf}.

\subsection{Diagrammatic representation}
\label{subsec:Diagram}

\begin{figure*}[htb]
\centering
\includegraphics[width=6.3cm, height=4cm]{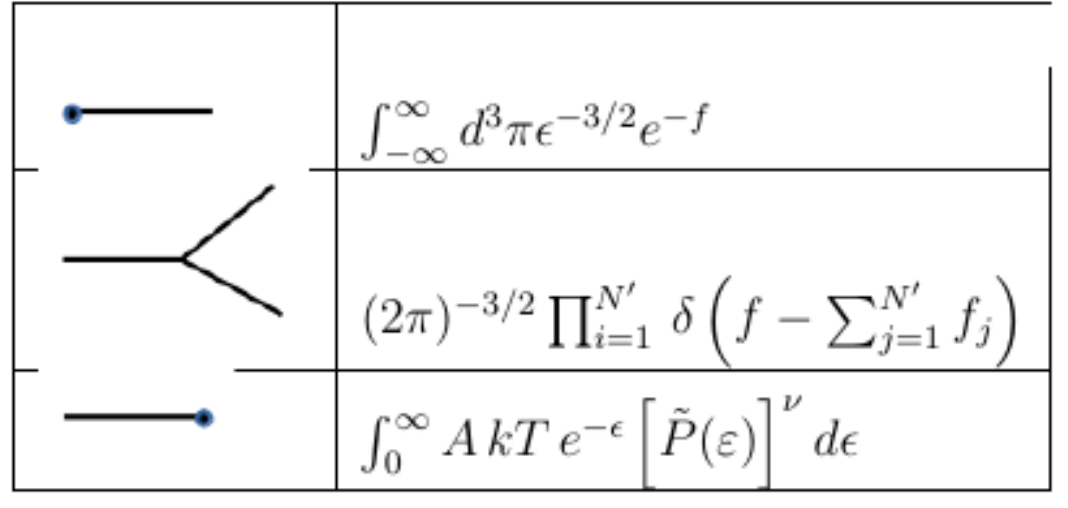} \hspace{1cm}
\includegraphics[width=4.5cm]{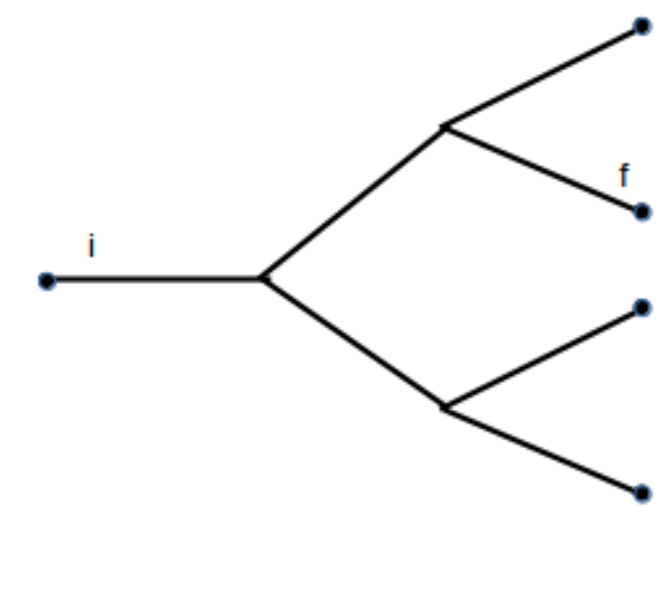}  
\caption{Left panel: Basic diagrams for thermofractals and their mathematical expressions. Right panel: Example of a tree graph representing different levels of a thermofractal.}
\label{fig:2}
\end{figure*}

Finally, let us mention that by using the formalism presented in Sec.~\ref{sec:Thermofractals}, it is possible to have a diagrammatic representation of the probability densities of thermofractals that can facilitate the calculation of the partition function and other relevant quantities~\cite{Deppman:2017fkq}. The basic diagrams are summarized in Fig.~\ref{fig:2}~(left). An example of diagram which gives the probability to find a constituent subsystem fractal $f$ at the third level of the initial fractal~$i$ is displayed in Fig.~\ref{fig:2}~(right) (see~\cite{Deppman:2017fkq} for further details). 

\section{Conclusions}
\label{sec:Conclusions}

We have introduced the non-extensive statistics in the form of Tsallis statistics of a quantum gas at finite temperature and chemical potential, and we have shown its application to study the EoS of QCD. In a second step we have investigated the structure of a thermodynamical system presenting fractal structure. It has been shown that the fractal structure in thermodynamics naturally leads to non-extensive statistics in the form of Tsallis. Using this formalism, we have introduced a diagrammatic formulation for practical calculations. Finally, based on the scale invariance of thermofractals, the Callan-Symanzik equation has been obtained. These results open the opportunity to develop a {\it 'field theoretical approach'} for thermofractals, leading to a possible theoretical understanding of the non-extensive properties of hadronic systems. This study will be addressed in details in~\cite{Deppman:inprep}.

\vspace{-0.5cm}
\begin{acknowledgement}
\section*{Acknowledgement}
A.D., D.P.M. and T.F. are partially supported by the Conselho Nacional de Desenvolvimento Cient\'{\i}fico e Tecnol\'ogico (CNPq-Brazil) and by Project INCT-FNA Proc. No. 464898/2014-5. T.F. thanks the partial support from Fundaç\~ao de Amparo \`a Pesquisa do Estado de S\~ao Paulo, FAPESP Grant No.~17/05660-0. The work of E.M. is supported by the Spanish MINEICO under Grants FPA2015-64041-C2-1-P and FIS2017-85053-C2-1-P, the Junta de Andaluc\'{\i}a under Grant FQM-225, the Basque Government under Grant IT979-16, and the Spanish Consolider Ingenio 2010 Programme CPAN (CSD2007-00042). The research of E.M. is also supported by the Ram\'on y Cajal Program of the Spanish MINEICO, and by the Universidad del Pa\'{\i}s Vasco UPV/EHU, Bilbao, Spain, as a Visiting Professor.
\end{acknowledgement}

%

\begin{thebibliography}{100}

\bibitem{Tsallis:1987eu}
C.~Tsallis, J. Statist. Phys. \textbf{52}, 479 (1988).

\bibitem{Tempesta:2011vc}
P.~Tempesta, Phys. Rev. \textbf{A84}, 021121 (2011).

\bibitem{Kalogeropoulos:2014mka}
N.~Kalogeropoulos, Int. J. Mod. Phys. \textbf{B28}, 1450162 (2014).

\bibitem{Borland:1998}
L.~Borland, Phys. Lett. \textbf{A245}, 67 (1998).

\bibitem{Wilk:2009nn}
G.~Wilk, Z.~Wlodarczyk, Phys. Rev. \textbf{C79}, 054903 (2009).

\bibitem{Deppman:2016fxs}
A.~Deppman, Phys. Rev. \textbf{D93}, 054001 (2016).

\bibitem{Megias:2015fra}
E.~Megias, D.P. Menezes, A.~Deppman, Physica \textbf{A421}, 15 (2015).

\bibitem{Hagedorn:1984hz}
R.~Hagedorn, Lect. Notes Phys. \textbf{221}, 53 (1985).

\bibitem{Menezes:2014wqa}
D.P. Menezes, A.~Deppman, E.~Megias, L.B. Castro, Eur. Phys. J. \textbf{A51},
  155 (2015).

\bibitem{Cleymans:1999st}
J.~Cleymans, K.~Redlich, Phys. Rev. \textbf{C60}, 054908 (1999).

\bibitem{Tawfik:2005qn}
A.~Tawfik, Nucl. Phys. \textbf{A764}, 387 (2006).

\bibitem{Deppman:2016prl}
A.~Deppman, E.~Megias, EPJ Web Conf. \textbf{141}, 01011 (2017).

\bibitem{Deppman:2017fkq}
A.~Deppman, T.~Frederico, E.~Megias, D.P. Menezes, Entropy \textbf{20}, 633
  (2018).

\bibitem{Deppman:2012us}
A.~Deppman, Physica \textbf{A391}, 6380 (2012).

\bibitem{Megias:2014tha}
E.~Megias, D.P. Menezes, A.~Deppman, EPJ Web Conf. \textbf{80}, 00040 (2014).

\bibitem{Deppman:2015cda}
A.~Deppman, E.~Megias, D.~Menezes, J. Phys. Conf. Ser. \textbf{607}, 012007
  (2015).

\bibitem{Deppman:2017igr}
A.~Deppman, Universe \textbf{3}, 62 (2017).

\bibitem{Deppman:inprep}
A.~Deppman, T.~Frederiro, E.~Megias, D.P. Menezes, in progress  (2018).

\bibitem{Ghosh:1998gw}
D.~Ghosh, A.~Deb, R.~Chattopadhyay, S.~Sarkar, A.K. Jafry, M.~Lahiri, S.~Das,
  K.~Purkait, B.~Biswas, J.~Roychoudhury, Phys. Rev. \textbf{C58}, 3553 (1998).

\bibitem{Deppman:2017rtf}
A.~Deppman, Adv. High Energy Phys. \textbf{2018}, 9141249 (2018).


\end{thebibliography}
%
%

\end{document}